\documentclass[preprint,showpacs,preprintnumbers,amsmath,amssymb]{revtex4}
\usepackage{booktabs}
\usepackage{mathrsfs}
\usepackage{epsfig}
\usepackage{graphicx}
\usepackage{dcolumn}
\usepackage{bm}
\usepackage{amsmath}
\usepackage{slashed}
\usepackage{multirow}

\let\jnfont=\rm
\def\NPB#1,{{\jnfont Nucl.\ Phys.\ B }{\bf #1},}
\def\PLB#1,{{\jnfont Phys.\ Lett.\ B }{\bf #1},}
\def\EPJC#1,{{\jnfont Eur.\ Phys.\ Jour.\ C }{\bf #1},}
\def\PRD#1,{{\jnfont Phys.\ Rev.\ D }{\bf #1},}
\def\PRL#1,{{\jnfont Phys.\ Rev.\ Lett.\ }{\bf #1},}
\def\MPLA#1,{{\jnfont Mod.\ Phys.\ Lett.\ A }{\bf #1},}
\def\JPG#1,{{\jnfont J.\ Phys.\ G}{\bf #1},}
\def\CTP#1,{{\jnfont Commun.\ Theor.\ Phys.\ }{\bf #1},}
\def\ZPC#1,{{\jnfont Z.\ Phys.\ C }{\bf #1},}
\def\JHEP#1,{{\jnfont JHEP \ }{\bf #1},}
\def\Rv{\not{\hbox{\kern-1pt $R$}}}
\def\p{\not{\hbox{\kern-3pt $p$}}}

\newcommand{\beq}{\begin{eqnarray}}
\newcommand{\eeq}{\end{eqnarray}}
\newcommand{\bpmatrix}{\begin{pmatrix}}
\newcommand{\epmatrix}{\end{pmatrix}}
\newcommand{\fr}{\frac}
\newcommand{\la}{\lambda}
\newcommand{\crn}{\nonumber \\}
\newcommand{\ep}{\epsilon}
\newcommand{\hc}{\text{ h.c}}

\newcommand{\diag}{\text{diag}}

\newcommand{\ba}{\begin{array}}
\newcommand{\ea}{\end{array}}

\newcommand{\be}{\begin{equation}}
\newcommand{\ee}{\end{equation}}

\begin{document}

\title{Higgs self-coupling in the MSSM and NMSSM after the LHC Run 1}

\author{Lei Wu$^1$\footnote{leiwu@physics.usyd.edu.au}, Jin Min Yang$^2$\footnote{jmyang@itp.ac.cn}, C.-P. Yuan$^3$\footnote{yuan@pa.msu.edu}, Mengchao Zhang$^2$\footnote{mczhang@itp.ac.cn}}
\affiliation{$^1$ ARC Centre of Excellence for Particle Physics at the Terascale, School of Physics, The University of Sydney, NSW 2006, Australia\\
$^2$ State Key Laboratory of Theoretical Physics, Institute of Theoretical Physics, Academia Sinica, Beijing 100190, China\\
$^3$ Department of Physics and Astronomy, Michigan State University, East Lansing, Michigan 48824, USA}%

\date{\today}

\begin{abstract}
Measuring the Higgs self-coupling is one of the crucial physics goals at the LHC Run-2 and 
other future colliders. In this work, we attempt to figure out the size of 
SUSY effects on the trilinear self-coupling of the 125 GeV Higgs boson in the MSSM and NMSSM 
after the LHC Run-1. Taking account of current experimental constraints, such as the Higgs data, 
flavor constraints, electroweak precision observables and dark matter detections, we obtain the
observations: (1) In the MSSM, the ratio of $\lambda^{MSSM}_{3h}/\lambda^{SM}_{3h}$ has been tightly 
constrained by the LHC data, which can be only slightly smaller than 1 and minimally reach 97\%; 
(2) In the NMSSM with $\lambda<0.7$, a sizable reduction of $\lambda^{NMSSM}_{3h_2}/\lambda^{SM}_{3h_2}$ 
can occur and minimally reach 10\% when the lightest CP-even Higgs boson mass $m_{h_1}$ is close to 
the SM-like Higgs boson $m_{h_2}$ due to the large mixing angle between the singlet and doublet Higgs 
bosons; (3) In the NMSSM with $\lambda>0.7$, a large enhancement or reduction 
$-1.1<\lambda^{NMSSM}_{3h_1}/\lambda^{SM}_{3h_1}<2$ can occur, which is accompanied by a sizable change 
of $h_1\tau^+\tau^-$ coupling. The future colliders, such as the HL-LHC and ILC, will have the capacity 
to test these large deviations in the NMSSM.
\end{abstract}

\maketitle

\section{Introduction}
\label{sec:intro}

With the discovery of the Higgs boson at the LHC \cite{atlas,cms}, much effort has been devoted to 
study its properties. So far, the measurements of its couplings and quantum numbers are compatible 
with the standard model (SM) predictions at 1-2$\sigma$ level. However, to ultimately understand 
its nature, we need to fully reconstruct the Higgs potential at the LHC and future $e^+e^-$ 
colliders \cite{higgs-review,hf}. The parameters in the Higgs potential determine the relations 
among the Higgs masses and self-couplings. Measuring these relations is therefore crucial for 
our understanding of the Higgs nature.

In the SM the tree-level Higgs potential is given by
\begin{equation}
V^{(0,SM)}=-\mu^2 (\phi^\dagger\phi)+\lambda(\phi^\dagger\phi)^2,\quad \phi=\frac{1}{\sqrt{2}}\left(0, v+h \right)^T
\end{equation}
which yields the following trilinear and quartic self-couplings
\begin{equation}
\lambda^{(0,SM)}_{hhh}=\frac{3m^2_{h}}{v},
\quad \lambda^{(0,SM)}_{hhhh}=\frac{3m^2_{h}}{v^2}\label{3h-4h}.
\end{equation}
Here $v=(\sqrt{2}G_F)^{-1/2}\simeq 246$ GeV is the
vacuum expectation value of the Higgs field and $m_h\simeq 125$ GeV is the Higgs boson mass. Within the SM, the trilinear Higgs coupling receives the dominant correction from the top quark loop, $\delta\lambda^{SM}_{hhh} \simeq m^4_t/(\pi^2 v^2 m^2_h)$ \cite{sm-self}, which reduces its tree-level value by about 10\%. Hence, the determination of the Higgs trilinear coupling $\lambda_{hhh}$ and quartic coupling $\lambda_{hhhh}$ can directly test the relation in Eq.~(\ref{3h-4h}) which is obtained from the minimization of the Higgs potential. At the LHC, the only way to measure the Higgs trilinear coupling is through the Higgs pair production, which is dominated by the gluon fusion mechanism and has a small cross section \cite{hh-sm}. However, in many new physics models, such as the minimal supersymmetric model (MSSM) and the next-to-minimal supersymmetric model (NMSSM), the Higgs pair production rate can be significantly altered by new particles and Higgs couplings \cite{hh-susy}. Among various decay channels of the Higgs pair, the $4b$ final state has the largest
fraction \cite{hh-mc1}, but the rare process $hh \to b\bar{b}\gamma\gamma$ is expected to have the most promising sensitivity due to the low backgrounds at the LHC \cite{hh-mc2}.
The recent applications of jet substructure and other techniques to Higgs pair production have been found to improve the sensitivity to the trilinear Higgs couplings in
$\tau^+\tau^-$ and $W^+W^-$ final states \cite{hh-mc3}. On the other hand, the measurement of the Higgs quartic coupling is more challenging due to a much smaller cross section
of triple Higgs production at the LHC.

The supersymmetric (SUSY) corrections to the Higgs self-coupling have two kinds of sources: one is the 
mixing between the Higgs bosons, and the other is radiative quantum effects. For the first kind, the 
authors in \cite{yoself} studied the Higgs self-coupling in some simplified SUSY models, 
while in \cite{james} the Higgs self-couplings in the MSSM and NMSSM (with a decoupled singlet boson) 
were investigated. Also, in \cite{barbieri} the authors studied the properties of the Higgs bosons
in the NMSSM with $\lambda>0.7$ (called $\lambda$-SUSY) and found a sizable enhancement in the Higgs self-coupling. 
For the second kind of SUSY corrections, the loop corrections to the Higgs self-couplings have been studied using the 
effective potential \cite{eff-mssm1,eff-mssm2,eff-mssm3,eff-mssm4,eff-mssm5,eff-nmssm0,eff-nmssm1,eff-nmssm2,eff-nmssm3}
or Feynman diagrammatic approach \cite{feyn-mssm1,feyn-mssm2,feyn-nmssm1} in the MSSM and NMSSM. 
Recently, the leading two-loop SUSY-QCD corrections from the top/stop sector
in the MSSM have been performed \cite{feyn-mssm3}. Since all these previous studies are limited to some simplified or 
special cases and the relevant experimental constraints are not fully considered, in this work we give a comprehensive 
study for the SM-like Higgs self-coupling in the MSSM and NMSSM with both $\lambda < 0.7$ and $\lambda > 0.7$ by 
considering all the relevant experimental constraints after the LHC Run-1.

The existing experimental data, both from low energy precision measurements and high energy direct searches, 
may have imposed important constraints on the Higgs self-coupling in SUSY models. For example, in the MSSM, 
the SM-like Higgs self-coupling is sensitive to the pseudo-scalar mass $m_A$ and $\tan\beta$, which could 
have been tightly constrained by the LHC direct search of a light non-SM Higgs boson 
\cite{cms-heavy,atlas-tautau,atlas-charged} as well as from precision $B$-physics \cite{bs}.
Furthermore, the measured mass of the SM-like Higgs boson requires rather heavy stops and/or large Higgs-stop 
trilinear couplings, and the direct searches for the stop pair production have also pushed stop masses above hundreds 
of GeV in the natural SUSY \cite{stop}. Since the Higgs self-couplings are sensitive to stop masses and Higgs-stop 
trilinear couplings, all these constraints should be taken into account.

The structure of this paper is organized as follows. In Section \ref{section2}, we will briefly describe the Higgs 
sectors of MSSM and NMSSM. In Section \ref{section3}, we perform a scan over the parameter space of each model and 
present the numerical results for the trilinear self-coupling of the SM-like Higgs. Finally, we draw our conclusions 
in Section \ref{section4}.

\section{Higgs trilinear self-couplings in MSSM and NMSSM \label{section2}}

\subsection{Higgs trilinear self-couplings in the MSSM}
In the MSSM there are two doublets of complex scalar fields with opposite hypercharges:
\begin{eqnarray}
H_u = \left(\begin{array}{c} H_u^+ \\ H_u^0 \end{array} \right), \qquad
H_d = \left(\begin{array}{c} H_d^0 \\ H_d^- \end{array} \right) \, .
\end{eqnarray}
The scalar Higgs potential consists of the $D$-terms and $F$-terms of the superpotential as well as the soft 
SUSY-breaking mass terms. Among them, the $D$-terms determine the
quartic Higgs interactions. The full tree-level Higgs potential is given by
\begin{eqnarray}
V^{(0,MSSM)} & = & m_1^2 |H_u|^2 + m_2^2 |H_d|^2 - B_\mu \epsilon_{\alpha\beta} (H_u^\alpha H_d^\beta
+ h.c.) \nonumber \\
& & +\frac{g^2+g'^2}{8} (|H_u|^2-|H_d|^2)^2 + \frac{g^2}{2}
|H_u^\dagger H_d|^2 \; ,
\label{mssmp}
\end{eqnarray}
where $\epsilon_{\alpha\beta}$ is the antisymmetric tensor and $m_{1,2}^2=m_{H_{u,d}}^2 + \mu^2$ with $m_{H_{u,d}}$ 
and $\mu$ denoting the soft SUSY-breaking masses and the higgsino mass, respectively. 
The parameters $m_{1,2}$ can be eliminated by the minimization condition of the Higgs potential, 
while the parameter $B_\mu$ is traded for the pseudoscalar mass $M_A$. The quartic Higgs couplings are fixed 
in terms of the $SU(2)\times U(1)$ gauge couplings $g$ and $g'$ in the MSSM.

After the electroweak symmetry breaking, the neutral components of the two Higgs fields $H_{u,d}^0$ develop 
vacuum expectation values (vevs) $v_{u,d}$ and can be decomposed into scalar and pseudoscalar components as
\begin{eqnarray}
{\rm Re} \, H_d^0 &=& (v_d+ H c_\alpha - h s_\alpha)/\sqrt{2}, \quad {\rm Im} \, H_d^0 = (G^0 c_\beta - A s_\beta)/\sqrt{2},\\
{\rm Re} \, H_u^0 &=& (v_u+ H s_\alpha + h c_\alpha)/\sqrt{2}, \quad {\rm Im} \, H_u^0 = (G^0 s_\beta + A c_\beta)/\sqrt{2}
\end{eqnarray}
where $h,H$ and $A$ are the neutral physical Higgs bosons and $G^0$ is the would-be Goldstone boson. The vevs are 
defined as $v_u = v s_\beta$ and $v_d = v c_\beta$ with $v\approx 246$ GeV (here and in the following we use the 
notation $c_x \equiv \cos x, s_x \equiv \sin x$).

Taking the third derivatives of $V^{(0,MSSM)}$ with respect to the physical Higgs fields yields the trilinear Higgs couplings. In the physical mass eigenstates, the neutral CP-even Higgs trilinear couplings at leading order are given by
\begin{eqnarray}\label{eq:mssm-3h}
\lambda^{(0,MSSM)}_{hhh} & = & \frac{3M_Z^2}{v}c_{2\alpha}s_{\alpha+\beta}, \quad
\lambda^{(0,MSSM)}_{Hhh}   =  \frac{M_Z^2}{v} [2s_{2\alpha}s_{\alpha+\beta} - c_{2\alpha}c_{\alpha+\beta}], \\
\lambda^{(0,MSSM)}_{HHH} & = & \frac{3M_Z^2}{v}c_{2\alpha}c_{\alpha+\beta}, \quad
\lambda^{(0,MSSM)}_{HHh}   =  -\frac{M_Z^2}{v} [2s_{2\alpha}c_{\alpha+\beta} + c_{2\alpha}s_{\alpha+\beta}].
\end{eqnarray}
In the MSSM, either the lighter scalar $h$ or the heavier scalar $H$ can be the SM-like Higgs boson. The latter interpretation occurs for low values of $M_A$ (between 100 and 120 GeV) with moderate values of $\tan\beta$ (about 10). In this case, $H$ has approximately SM-like properties, while the other four Higgs bosons of the MSSM would be rather light and have a mass of order 100 GeV or even below. A dedicated scan for this region of parameter space has been performed in \cite{arbey} and 
it was found that this scenario can be excluded by recasting the LHC search for $H/A \to \tau^+\tau^-$ \cite{cms-heavy}. 
In addition, the latest ATLAS limits from $H^\pm$ searches have also excluded such a possibility \cite{atlas-charged}. Thus, in this work, we will only study the case that $h$ is the SM-like Higgs boson in the MSSM. According to Appelquist-Carazzone decoupling theorem, the MSSM must go back to the SM in the decoupling limit. In the tree-level Higgs sector, we can make this limit by setting $m_A \to \infty$, which gives $\alpha \to \beta -\frac{\pi}{2}$.
Applying this relation to the first identity in Eq.~(\ref{eq:mssm-3h}) yields
\begin{equation}
\lambda^{(0,MSSM)}_{hhh}  \simeq  \frac{3M_Z^2}{v}c^2_{2\beta} \simeq \frac{3m^{2}_{h,(0,MSSM)}}{v},
\end{equation}
where $m_{h,(0,MSSM)}\simeq M_Zc_{2\beta}$ is the lighter CP-even Higgs mass at tree-level.  
This demonstrates that the lighter Higgs boson $h$ in the MSSM almost behaves like the SM Higgs boson in the decoupling 
limit (even when the loop corrections are included) \cite{feyn-mssm1,feyn-mssm2}.

\subsection{Higgs trilinear self-couplings in the NMSSM}
After the discovery of a 125 GeV Higgs boson, the NMSSM  \cite{NMSSM-Review} seems to be more favored than the MSSM because it can naturally give such a Higgs boson without very heavy top-squarks \cite{cao-nmssm}. More importantly, this model can solve the $\mu$-problem: after the singlet field develops a vev $\langle S \rangle=v_s/\sqrt{2}$, an effective $\mu$-term ($\mu_{eff}=\lambda v_s/\sqrt{2}$) is dynamically generated. Due to the contribution of the singlet scalar field $S$, the full tree-level Higgs potential can be written as
\beq
V^{(0,NMSSM)} &=& (|\la S|^2 + m_{H_u}^2)H_{u}^\dagger H_{u} + (|\la S|^2 + m_{H_d}^2)H_{d}^\dagger H_{d} +m_S^2 |S|^2 \crn
&& + \fr18 (g_2^2+g_1^{2})( H_{u}^\dagger H_{u}-H_{d}^\dagger H_{d} )^2
+\fr12g_2^2|H_{u}^\dagger H_{d}|^2\crn
&&   + |\ep^{\alpha\beta} \la  H^{\alpha}_{u}  H^{\beta}_{d} + \kappa S^2 |^2+
\big[\ep^{\alpha\beta}\la A_\la H^{\alpha}_{u}  H^{\beta}_{d} S  +\fr13 \kappa
A_{\kappa} S^3+\hc \big] \,,
\label{nmssmp}
\eeq
where $\kappa$ and $\lambda$ are dimensionless parameters, and $A_{\lambda}$ and $A_{\kappa}$ are the corresponding trilinear soft breaking parameters. To clearly show the properties of the Higgs sector, we can expand the neutral scalar fields around the vevs as \cite{NMSSM-Review}
\begin{eqnarray}
{\rm Re} \, H_d^0 &=& (v_d- H \sin\beta +h \cos\beta)/\sqrt{2}, \quad {\rm Im}\, H_d^0= (P \sin \beta + G^0 \cos \beta)/\sqrt{2}, \nonumber \\
{\rm Re} \, H_u^0 &=& (v_u+ H \cos\beta +h \sin\beta)/\sqrt{2}, \quad {\rm Im}\, H_u^0= (P \cos \beta - G^0 \sin \beta)/\sqrt{2}, \nonumber \\
{\rm Re} \, S &=& (v_s+s)/\sqrt{2}, \quad \quad \quad \quad \quad \quad \quad \quad
{\rm Im}\, S= P_S/\sqrt{2}.
\label{nmssm-higgs}
\end{eqnarray}
Substituting Eq.~(\ref{nmssm-higgs}) into Eq.~(\ref{nmssmp}), we obtain the mass matrix squared $M_S^2$ for the neutral 
CP-even Higgs bosons and the trilinear Higgs self-interactions as 
\begin{align} V^{(0,NMSSM)}_{\rm CP-even} =
 \fr 12 \bpmatrix H,h,s \epmatrix  {M_S^2}
\bpmatrix  H\\h\\s \epmatrix   +\la_{h_\alpha h_\beta h_\gamma} h_\alpha h_\beta h_\gamma \,,
\label{eq:vhpot}
\end{align}
with $h_{\alpha,\beta,\gamma}=H,h,s$. The tree-level $M_{S\,ij}^2$ are given by \cite{roman}
\begin{eqnarray}\label{mass1}
M_{S\,11}^2&=& M_A^2 + (M_Z^2- \frac{1}{2} \lambda^2 v^2) \sin^2 2 \beta,\\
M_{S\,12}^2&=&- \frac{1}{2} (M_Z^2- \frac{1}{2} \lambda^2 v^2) \sin 4 \beta, \\
M_{S\,13}^2&=&-\sqrt{2}\lambda v\mu x \cot2\beta,\\
M_{S\,22}^2&=&M_Z^2\cos^2 2 \beta + \frac{1}{2} \lambda^2 v^2 \sin^2 2 \beta,\label{sm-like}\\
M_{S\,23}^2&=&\sqrt{2}\lambda v\mu (1-x),\\
M_{S\,33}^2&=&4\frac{\kappa^2}{\lambda^2}\mu^2+\frac{\kappa}{\lambda}A_{\kappa}\mu+\frac{\lambda^2v^2}{2}x-\frac{\kappa\lambda}{2}v^2\sin2\beta,\label{mass6}
\end{eqnarray}
with
\begin{eqnarray}
M_A^2 &=& \frac{\lambda v_s}{\sin 2 \beta} \left( \sqrt{2} A_{\lambda} + \kappa v_s \right), \quad x = \frac{1}{2\mu}(A_\lambda+2\frac{\kappa}{\lambda}\mu).
\end{eqnarray}
Here, it should be mentioned that the mass parameter $M_A$ in the NMSSM becomes the mass of the heavy pseudoscalar 
Higgs boson only in the MSSM limit ($\lambda, \kappa \to 0$ with the ratio $\kappa/\lambda$ fixed). 
In the NMSSM, $M_A$ can be traded by the soft parameter $A_\lambda$.

The CP-even Higgs mass eigenstates $h_i$ ($i=1,2,3$) can be obtained by diagonalizing ${M_S^2}$ with a rotation matrix ${\cal O}$
\beq
 h_i =  {\cal O}_{i\alpha}h_{\alpha},~~ (h_\alpha=H,h,s), \quad \diag(m_{h_1}^2,m_{h_2}^2,m_{h_3}^2)= {\cal O} {M_S^2} {\cal O}^T \label{eq:rotationS} \;
\eeq
with ${\cal O}_{i\alpha}$ being the elements of the rotation matrix satisfying the sum
rules
\begin{eqnarray}
{\cal O}^2_{1\alpha} + {\cal O}^2_{2\alpha} + {\cal O}^2_{3\alpha}&=&1.
\end{eqnarray}
The mass eigenstates $h_{i}$ are aligned by the masses $m_{h_1} \le m_{h_2} \le m_{h_3}$. The singlet or non-SM doublet components in a physical Higgs boson $h_i$ is determined by the rotation matrix elements ${\cal O}_{i(H,s)}$. With Eq.~(\ref{eq:rotationS}), the corresponding tree-level trilinear Higgs couplings in the mass eigenstates $h_i$ are given by
\beq
\lambda^{(0,NMSSM)}_{h_i h_j h_k} = {\cal O}_{i\alpha} {\cal O}_{j\beta} {\cal O}_{k\gamma} \lambda_{h_\alpha h_\beta h_\gamma} .\label{3h}
\eeq

In the NMSSM, we take $h_1$ or $h_2$ as the SM-like Higgs boson when $|{\cal O}_{(1,2)h}|^2 \geq 0.5$. In general, due to 
the introduction of the singlet $s$ and its couplings to the MSSM Higgs sector, the mass of the SM-like Higgs boson 
$M^2_{S_{22}}$ can be lifted by the extra large $\lambda$-term at tree level, as shown in Eq.~(\ref{sm-like}). The value 
of $\lambda$ 
at the weak scale is upper bounded by $0.7$ in order for the NMSSM to remain perturbative up to the GUT 
scale \cite{NMSSM-Review,roman}. 
Whereas, the case of $\lambda > 0.7$ (dubbed as $\lambda$-SUSY model) is still of interest because it can suppress 
the sensitivity of the Higgs mass with respect to changes of the soft SUSY-breaking masses and keep the fine tuning at 
a moderate level even for stop masses up to 1 TeV. So, in our study, we consider both $\lambda < 0.7$ and $\lambda>0.7$ 
cases:
\begin{itemize}
  \item For $\lambda < 0.7$, in addition to the tree-level $\lambda$ contribution, the mixture of the singlet $s$ with the 
MSSM Higgs $h$ and $H$, as shown in Eq.~(\ref{mass1}), in particular with $h$, could further modify the SM-like Higgs mass. 
If $H$ is decoupled, when $M^2_{S_{22}} > M^2_{S_{33}}$, the mass eigenvalues for the SM-like Higgs boson $m_{h_2}$ is pushed up 
by the positive mixing effect after the diagonalization of the $h-s$ mass matrix. However, when $M^2_{S_{22}} < M^2_{S_{33}}$, 
the mass eigenvalues for the SM-like Higgs boson $m_{h_1}$ is pulled down by the negative mixing effect. Without the negative
mixing effect, the maximal tree-level SM-like Higgs mass $m_{h_1}$ can only reach to about 110 GeV \cite{ellwanger,agashe}, 
which needs a sizable loop correction from the stop sector to obtain a 125 GeV Higgs boson like in the MSSM. In this sense, 
$h_2$ being the observed SM-like Higgs boson may be more natural than $h_1$ in the NMSSM. So, we choose the next-to-lightest
CP-even Higgs boson $h_2$ as the SM-like Higgs boson in the following discussions for the NMSSM with $\lambda<0.7$. 
Note that the lightest Higgs boson $h_1$ in this case is predominantly singlet-like and its mass can be as light as 
about 20 GeV in our scanned samples. Consequently, the SM-like Higgs boson $h_2$ can decay into a pair of light scalars 
$h_1$ and hence the $\gamma\gamma$ and $ZZ^*$ signal rates are suppressed. In order to be consistent with the LHC Higgs 
data, the branching ratio of $h_2 \to h_1 h_1$ was found to be less than about 30\% ~\cite{yang} and may be tested through 
$h_1 h_1 \to b\bar{b}\mu^+\mu^-$ production channel at the 14 TeV LHC \cite{bbmumu}. However, due to our interests in the 
large mixing region, we only display the results with $m_{h_1} > m_{h_2}/2$ in the following analysis. We will also decouple 
the heaviest CP-even Higgs boson $h_3$ by requiring $m_{h_3}>1$ TeV and focus on the singlet-doublet system. On the other 
hand, as mentioned above, if $h_1$ is the SM-like Higgs boson, the value of $\lambda$ tends to be large in order to 
maximize the tree-level Higgs mass (so in our study for $\lambda$-SUSY with  $\lambda>0.7$, we will choose 
$h_1$ as the SM-like Higgs boson). This feature may lead to a sizable change in Higgs self-coupling when doublet-singlet 
mixing effect is large. We checked this possibility and found that the ratio $\lambda^{NMSSM}_{3h_1}/\lambda^{SM}_{3h}$ can 
vary from 0.29 to 1.17 for our samples with $\lambda<0.7$. 

  \item For $\lambda>0.7$, we choose the lightest CP even Higgs boson $h_1$ as the SM-like Higgs boson. The reason is 
that for $\lambda>0.7$ the tree-level Higgs mass will be significantly lifted. If $h_2$ is assumed as the SM-like Higgs 
boson, the large $\lambda$-term in $M^2_{S_{22}}$ and the positive doublet-singlet mixing effect can readily make the SM-like Higgs mass $m_{h_2}$ exceed 125 GeV. Thus, such a choice is strongly disfavored by the LHC observed Higgs mass \cite{agashe} and will not be further studied in our work. If $h_1$ is the SM-like Higgs boson, as pointed before, the cancelation between 
the tree-level $\lambda$-term and the negative doublet-singlet mixing effect can easily yield a 125 GeV SM-like Higgs boson.
However, $\lambda$ can not be too large, because the large trilinear and quartic couplings of singlet scalar field $s$ will largely contribute to the scattering amplitude $ss \to ss$. In the high energy limit, the unitarity condition requires $|\lambda| \leq 3$ and $|\kappa| \leq 3$. Furthermore, if combined with the dark matter relic abundance constraint, the above-mentioned unitarity bound can set a generic upper bound 20 TeV for the heavy Higgs masses \cite{unitarity-nmssm}. The requirement of perturbativity up to the cut-off scale $\Lambda$, i.e., $\lambda(\Lambda) \lesssim 2\pi$ and $\kappa(\Lambda) \lesssim 2\pi$, will set upper bounds on $\lambda$ and $\kappa$ at weak scale. In this study, we assume the new unknown strong dynamics, for restoring the unitary of scattering processes, appears at some scale $\Lambda$ above 10 TeV, and require $\lambda^2 + \kappa^2 \lesssim 4.2$ \cite{roman,cao-lambda}.

\end{itemize}
Before closing this section, we note that since both Higgs boson masses and Higgs self-interactions arise from the Higgs potential, one has to adopt the same method to calculate them up to the same order for comparing the self-couplings in SUSY and SM. Although the most accurate evaluation of Higgs mass is up to three-loop level in the MSSM \cite{higgsmass}, there is no corresponding result for the Higgs self-couplings. Hence, in our study, we derive the Higgs mass and self-couplings by following the effective potential approaches in the MSSM \cite{eff-mssm1} and NMSSM \cite{eff-nmssm2}, where the explicit dominant one- and two-loop corrections to the effective potential are presented. Then we coded those expressions in our numerical calculation. On the other hand, since the tree-level mixing effects in the Higgs sector are usually dominant over the loop corrections in the NMSSM, we will focus on the doublet-singlet mixing effects for the NMSSM results.

\section{Numerical calculations and results \label{section3}}
\subsection{Scan over the parameter space}

In our numerical calculations, we take the input parameters of the SM as \cite{pdg}
\begin{eqnarray}
&&m_t=173.5{\rm ~GeV}, \quad m_{W}=80.385{\rm ~GeV}, \quad m_{Z}=91.1876 {\rm ~GeV},\nonumber\\
&&m^{\overline{MS}}_{b}(m^{\overline{MS}}_{b})=4.18{\rm ~GeV}, \quad \alpha_s(M_Z)=0.1184, \quad \alpha(m_Z)^{-1}=128.962
\end{eqnarray}
We use \textsf{NMSSMTools-4.4.1} \cite{NMSSMTOOLS} to perform a random scan over the parameter space. Note that for any given value of $\mu_{eff}$, the phenomenology of the NMSSM is identical to the MSSM in the limit $\lambda, \kappa \to 0$ with the ratio $\kappa/\lambda$ fixed ($A_{\kappa}$ should be negative and satisfy $|A_\kappa| < 4\kappa \mu/\lambda$ to guarantee the squared mass of the singlet scalar to be positive) \cite{NMSSM-Review}. In our scan for the MSSM, we take $\lambda = \kappa = 10^{-7}$ and $A_\kappa = -10$ GeV.  The validity of such a method has been justified by the authors of the \textsf{NMSSMTools} \cite{NMSSMTOOLS} and our previous calculations \cite{cao-nmssm}. We also numerically checked our MSSM results of $m_h$ by using the codes \textsf{FeynHiggs} \cite{feynhiggs}, \textsf{SOFTSUSY} \cite{softsusy} and \textsf{SuSpect} \cite{suspect}, and found the results to agree with that given by the \textsf{NMSSMTools} within about 1\% level when $m_h \sim 125$ GeV. So it is feasible to use the package \textsf{NMSSMTools} in the MSSM limit to study the phenomenology of the MSSM. For simplicity, we decouple the sleptons, gauginos and the first two generations of squarks by fixing the corresponding soft mass parameters at 2 TeV. We also set $M_{Q3}=M_{D3}=M_{U3}$ and $A_t =A_b$ for the third generation of squarks. The lower limit of $\tan\beta$ in the MSSM is taken as 5, which is inspired by the recent LHC Higgs results \cite{arbey}. The parameter ranges in our scan are chosen as the following:
\begin{itemize}
\item[(a)] For the MSSM,
\begin{eqnarray}
&&5 \le \tan\beta \le 60, ~ 0.2 {~\rm TeV} \le \ M_A \le 1 {~\rm TeV},
~|\mu| \le 1 {~\rm TeV}, \nonumber \\
&&
0.1 {~\rm TeV} \le M_{Q3} \le 2.5 {~\rm TeV}, ~ |A_t| \le 3M_{Q3}.
\end{eqnarray}
\item[(b)] For the NMSSM with $\lambda < 0.7$,
\begin{eqnarray}
&&0 \le \lambda \le 0.7, ~ |\kappa| \le 0.7,
 ~0.2 {~\rm TeV} \le A_\lambda \le 1 {~\rm TeV},  \nonumber\\
&&
|A_\kappa| \le 1 {~\rm TeV}, ~1 \le \tan\beta \le 20,  ~~0.1 {~\rm TeV} \le \mu \le 1 {~\rm TeV}, \nonumber\\
&& 0.1 {~\rm TeV} \le M_{Q3} \le 1 {~\rm TeV}, ~~|A_t| \le 3M_{Q3}.
\end{eqnarray}
\item[(c)] For the NMSSM with $\lambda > 0.7$,
\begin{eqnarray}
&&0.7 < \lambda \le 2, ~ 0< \kappa \le 2,
 ~0.2 {~\rm TeV} \le A_\lambda \le 1 {~\rm TeV},  \nonumber\\
&&
|A_\kappa| \le 1 {~\rm TeV},~1 \le \tan\beta \le 20,  ~~0.1 {~\rm TeV} \le \mu \le 1 {~\rm TeV}, \nonumber\\
&& 0.1 {~\rm TeV} \le M_{Q3} \le 1 {~\rm TeV}, ~~|A_t| \le 3M_{Q3}.
\end{eqnarray}
\end{itemize}
In our scan, we consider the following constraints:
\begin{itemize}
\item[(1)] We require the SM-like Higgs mass in the range of 123-127 GeV and consider the exclusion limits (at the 95\% confidence level) from LEP, Tevatron and LHC in Higgs searches with \textsf{HiggsBounds-4.2.0} \cite{higgsbounds}. We also perform the Higgs data fit by calculating $\chi^{2}$ of the Higgs couplings with the public package \textsf{HiggsSignals-1.3.0} \cite{higgssignals} and require our samples to be consistent with Higgs data at $2\sigma$ level. We choose the SLHA input choice of HiggsBounds/HiggsSignals, where the effective Higgs couplings are only used to calculate the Higgs production cross section ratios. The Higgs decay branching ratios are taken directly from the corresponding decay blocks in the SLHA file generated by the \textsf{NMSSMTools}.
\item[(2)] We require one-loop SUSY predictions for $B$-physics observables to satisfy the 2$\sigma$ bounds as encoded in \textsf{NMSSMTools}, which include $B\to X_s\gamma$, $B_s\to \mu^+\mu^-$, $B_d\to X_s\mu^+\mu^-$ and $ B^+\to \tau^+\nu$. Theoretical uncertainties in B-physics observables are taken into account as implemented in \textsf{NMSSMTools}.
\item[(3)] We require the one-loop SUSY predictions for the precision electroweak observables such as $\rho_l$, $\sin^2 \theta_{\rm eff}^l$, $m_W$ and $R_b$ \cite{rb} to be
within the $2\sigma$ ranges of the experimental values.
\item[(4)] We require the thermal relic density of the lightest neutralino (as the dark matter candidate) is below the 2$\sigma$ upper bound of the Planck value \cite{planck}
 and the spin-independent neutralino-proton scattering cross section satisfy the direct detection bound from LUX at 90\% confidence level \cite{lux}.
\item[(5)] We also consider the theoretical constraints from the stability of the Higgs potential as encoded in the \textsf{NMSSMTools}.
\end{itemize}

\subsection{Results for the MSSM}
\begin{figure}[ht]
\centering
\includegraphics[width=3.5in]{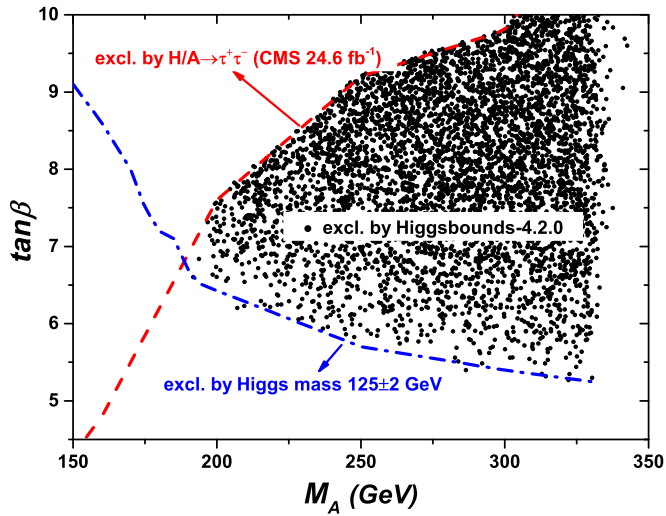}\vspace{-0.5cm}
\caption{Excluded region in the $\tan\beta$ versus $M_A$ plane of the MSSM.}
\label{mssm-low}
\end{figure}
Precision measurements of the Higgs boson properties (its mass and couplings to other particles) at the LHC provide relevant constraints on possible weak-scale extensions of the SM. In the usual context of the MSSM, these constraints suggest that all the additional non-SM-like Higgs bosons should be heavy. In Fig.~\ref{mssm-low}, we show that in the decoupling limit of the MSSM, the region with $m_A < 330$ GeV has been excluded by various constraints. Similar result has been recently pointed out in \cite{choudhury}. The lower part of Fig.~\ref{mssm-low} shows that the value of $\tan\beta$ can not be too small due to the requirement of 125-127 GeV Higgs mass in our scan. For a small $\tan\beta$, heavy stops ($\gtrsim 10$ TeV) or a large mixing parameter $A_t$ is needed to produce a large positive correction to the Higgs mass, which, however, will easily lead to vacuum instability and large uncertainty in the numerical calculation. So, we focus on $m_{\tilde{t}_1}<2.5$ TeV region in our calculations. The additional constraints on the MSSM parameters imposed by the Higgs data are obtained by using the \textsf{HiggsBounds-4.2.0} package. The key algorithm of \textsf{HiggsBounds} can be described in two steps. Firstly, the \textsf{HiggsBounds} uses the expected experimental limits from LEP, Tevatron and LHC to determine which decay channel has the highest statistical sensitivity. Secondly, for this particular channel, the theory prediction is compared to the observed experimental limits to conclude whether this sample is allowed or excluded at 95\% CL. With the \textsf{HiggsBounds}, we find that most of samples (black bullets) with $180 \lesssim M_A \lesssim 330$ GeV and $5 \lesssim \tan\beta \lesssim 10$ have been excluded. Particularly, the latest CMS result of searching for $H/A \to \tau^+\tau^-$ has excluded most parameter space with a low $M_A$ and low to moderate values of $\tan\beta$, as shown in the upper left corner of Fig.~\ref{mssm-low}. We have checked that the current low energy constraints from $B_s \to \mu^+\mu^-$ and $B_s \to X_s \gamma$ are weaker than $H/A \to \tau^+\tau^-$ for our interested region (low to moderate values of $\tan \beta$). We note that the supersymmetric loop corrections generally lead to a contribution of the order of a few percent of the SM value. Hence, the suppressions in Higgs signal strengthes $\mu_{\gamma\gamma}$ and $\mu_{VV^*}$ are mostly governed by the increase of the width of the lightest CP-even Higgs decay into bottom quarks and tau leptons at low values of $m_A$.

Fig.~\ref{mssm-low} suggests that all the additional non-SM-like Higgs bosons should be heavy, with masses larger than about 330 GeV. This is the commonly discussed decoupling limit of the MSSM. However, as discussed in \cite{align-2}, it is also possible to have the MSSM parameter conditions for ¡°alignment independent of decoupling¡±, where the lightest
CP-even Higgs boson has SM-like tree-level couplings to fermions and gauge bosons, independently of the non-standard Higgs boson masses. In the alignment region, $\sin(\alpha-\beta) \sim 1$ and the bounds on the heavy Higgs bosons that arise from the measurements of $h \to VV$ may be relaxed. Such alignment conditions are associated with very SM-like
$ht\bar{t}$ coupling and tend to be restricted to values of $\tan\beta$ of order 10 or larger within the MSSM \cite{align-1, align-2,align-3}. As will be shown below, for such a large value of $\tan\beta$, the supersymmetric contributions to the SM-like Higgs self-coupling are found to be small (at a percent level). Hence, even in the alignment condition, we cannot
expect a large deviation of the Higgs self-coupling with the SM value.

\begin{figure}[ht]
\centering
\includegraphics[width=5in]{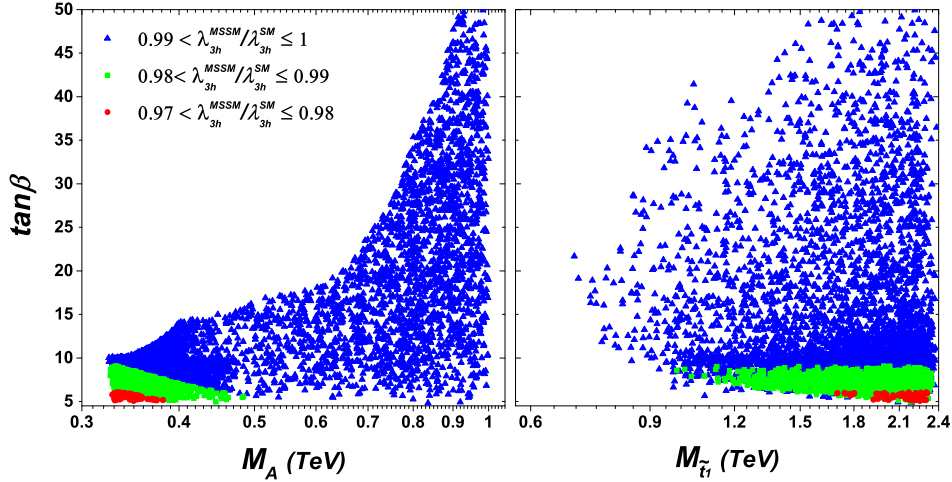}\vspace{-0.2cm}
\caption{Scatter plots of the samples surviving all the experimental constraints, projected on the planes of $\tan\beta$ versus $m_{A}$ and $m_{\tilde{t}_{1}}$.}
\label{mssm-3h}
\end{figure}
In Fig.~\ref{mssm-3h} we project the samples surviving all the experimental constraints on the planes of $m_{A}$ and $m_{\tilde{t}_{1}}$ versus $\tan\beta$. The ratio $\lambda^{MSSM}_{3h}/\lambda^{SM}_{3h}$ is always smaller than 1 because of the 
negative MSSM corrections to Higgs self-coupling \cite{feyn-mssm1}. From the left panel it can be seen that in most part of the allowed parameter space (the blue triangles), the values of $\lambda^{MSSM}_{3h}/\lambda^{SM}_{3h}$ (we use $3h$ to denote $hhh$) are larger than 0.99. When both $\tan\beta$ and $m_A$ become small, $\lambda^{MSSM}_{3h}/\lambda^{SM}_{3h}$ gets smaller, which can minimally reach about 0.97 for our samples. The reasons for $\lambda^{MSSM}_{3h}$ being so close to $\lambda^{SM}_{3h}$ are the following: (1) The dominant MSSM contributions to $\lambda^{MSSM}_{3h}/\lambda^{SM}_{3h}$ are from the stop loops, while the Higgs coupling with the stops is proportional to $1/\sin\beta$. So it leads to an overall enhancement factor $1/\sin^3\beta$ in the corrections when $\tan\beta$ is small. However, such a region is obviously not favored by the measured Higgs mass, which needs a large $\tan\beta$ to enhance the Higgs mass; (2) A light $m_{A}$ causes a large mixing between two CP-even Higgs bosons and can sizably change the Higgs couplings with the SM fermions. But as mentioned before, $m_A$ should be heavier than about 330 GeV to satisfy the experimental constraints in our scan. For the right panel it should be mentioned that since the small values of $\lambda^{MSSM}_{3h}/\lambda^{SM}_{3h}$ occur in the small $\tan\beta$ region, heavy stops are usually needed to enhance the Higgs mass through loop corrections.

\begin{figure}[ht]
\centering
\includegraphics[width=5in]{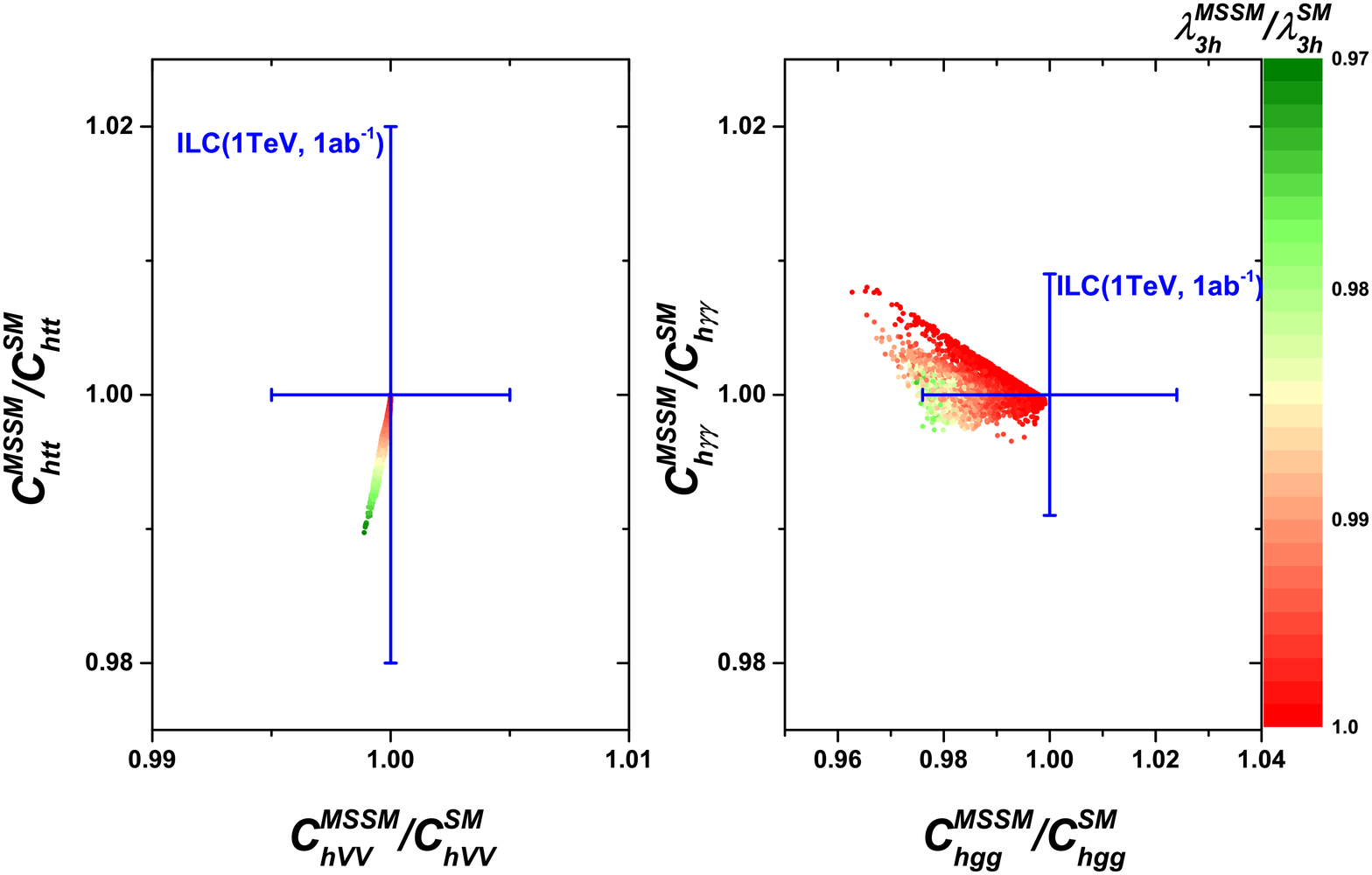}\vspace{-0.2cm}
\caption{Same as Fig.\ref{mssm-3h}, but showing the Higgs couplings. The ILC (1 TeV, 1 ab$^{-1}$) sensitivities to the alteration of the couplings \cite{snowmass} are also ploted
(the regions between the bars give too small alterations to be detectable at ILC).}
\label{mssm-couplings}
\end{figure}
In Fig.~\ref{mssm-couplings}, we show the MSSM Higgs couplings in comparison with the SM predictions. The ILC (1 TeV, 1 ab$^{-1}$) sensitivities to the alteration of the couplings
\cite{snowmass} are also ploted, where the regions between the bars give too small alterations to be detectable at ILC. The HL-LHC (14 TeV, 3 ab$^{-1}$) sensitivities are much worse than ILC and are not shown here. From Fig.~\ref{mssm-couplings}, we have the following observations: (1) In the MSSM, the Higgs gauge couplings and top-Higgs couplings are respectively changed by the factors $\sin(\beta-\alpha)$ and $\cos\alpha/\sin\beta$, but the ratios $C^{MSSM}_{hVV}/C^{SM}_{hVV}$ and $C^{MSSM}_{ht\bar{t}}/C^{SM}_{ht\bar{t}}$
for our survived samples are very close to unity due to $m_A \gg m_Z$. So, even if these couplings can be measured at percent level at ILC, it can not constrain the MSSM parameter space with a small $\lambda^{MSSM}_{3h}/\lambda^{SM}_{3h}$; (2) The ratio  $C^{MSSM}_{hgg}/C^{SM}_{hgg}$ is always smaller than one for our samples because a large mixing between the stops or heavy stops needed by the Higgs mass interferes destructively with the top loop and $C^{MSSM}_{hgg}$ is suppressed, while $C^{MSSM}_{h\gamma\gamma}/C^{SM}_{h\gamma\gamma}$ can be greater than one due to the constructive contribution with the $W$ loop. Although both couplings only slightly deviate from the SM predictions, the high precision measurements on $C_{gg,\gamma\gamma}$ at the ILC will be able to exclude some part of the parameter space with $\lambda^{MSSM}_{3h}/\lambda^{SM}_{3h}>0.977$.

\subsection{Results for the NMSSM ($\lambda<0.7$)}
\begin{figure}[ht]
\centering
\includegraphics[width=5in]{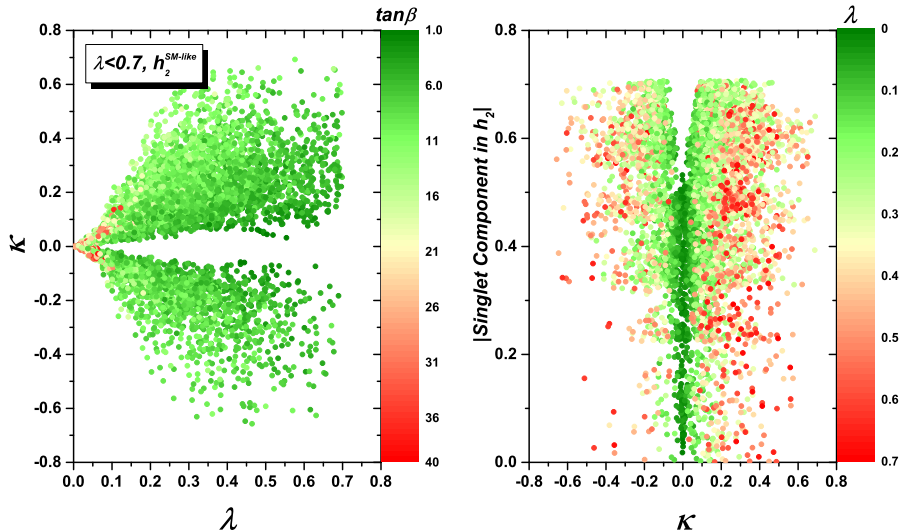}\vspace{-0.2cm}
\caption{The NMSSM ($\lambda<0.7$) samples surviving all the experimental constraints, projected on the planes of $\lambda$ versus $\kappa$ and the singlet component
$|{\cal O}_{2s}|$ in the SM-like Higgs boson $h_2$.}
\label{nmssm-pars}
\end{figure}
In Fig.~\ref{nmssm-pars}, we project the NMSSM ($\lambda<0.7$) samples allowed by the constraints (1)-(5) on the planes of $\lambda$ versus $\kappa$ and $\kappa$ versus the singlet
component $|{\cal O}_{2s}|$ in the SM-like Higgs boson $h_2$. Here we take $2m_{h_1}>m_{h_2}$ 
so that $h_2$ will not decay into a pair of $h_1$. We can see that when $\lambda$ and $\kappa$ approach to zero, $\tan\beta$ has to be large in order to enhance the Higgs mass. While in small $\tan\beta \lesssim 10$ region, the values of $\lambda$ for most samples are larger (in magnitude) than $\kappa$. Since the singlet component $|{\cal O}_{2s}|$ in the SM-like Higgs boson $h_2$ can potentially affect the Higgs self-coupling in Eq.~(\ref{3h}), we also show the dependence of $|{\cal O}_{2s}|$.
Since the singlet component $|{\cal O}_{2s}|$ in the SM-like Higgs boson $h_2$ can potentially affect the Higgs self-coupling in Eq.~(\ref{3h}), we also show the dependence of $|{\cal O}_{2s}|$ on $\lambda$ and $\kappa$ for the NMSSM 
(with $\lambda<0.7$) samples which survive all the experimental constraints. It can be seen that $|{\cal O}_{2s}|$ deceases if both $\lambda$ and $\kappa$ go to zero. However, if $\lambda \gg |\kappa| \sim 0$, which usually happens for small $\tan\beta \lesssim 10$, $|{\cal O}_{2s}|$ can be sizeable and lead to a large deviation from the SM prediction in triple Higgs boson coupling. For example, for $\lambda=0.017$ and $\kappa=-0.0038$, ${\cal O}_{2s} \sim 0.215$ and $\lambda^{NMSSM}_{3h_2}/\lambda^{SM}_{3h} \sim 0.7$. The reason is that in the singlet-doublet system, after decoupling $h_3$ by requiring $M_{h_3} >1$ TeV, $|{\cal O}_{2s}|$ is proportional to $\sin\theta$, where the mixing angle $\theta$ that determines the mixture of the singlet and the SM-like Higgs states can be approximately expressed as $\tan2\theta \sim 2 M_{23}^2 / (M_{22}^2 - M_{33}^2)$ \cite{roman,agashe}. Here, $M^2_{ij}$ are the CP-even Higgs mass matrix elements listed in Eqs.~(\ref{mass1}-\ref{mass6}). In case that $\lambda$ is not too small, a large $\theta$ can happen when $M_{22}^2 - M_{33}^2 \sim M_{23}^2$. This leads to a large singlet component in the SM-like Higgs state and hence a sizeable modification of the Higgs trilinear couplings.

\begin{figure}[ht]
\centering
\includegraphics[width=5in]{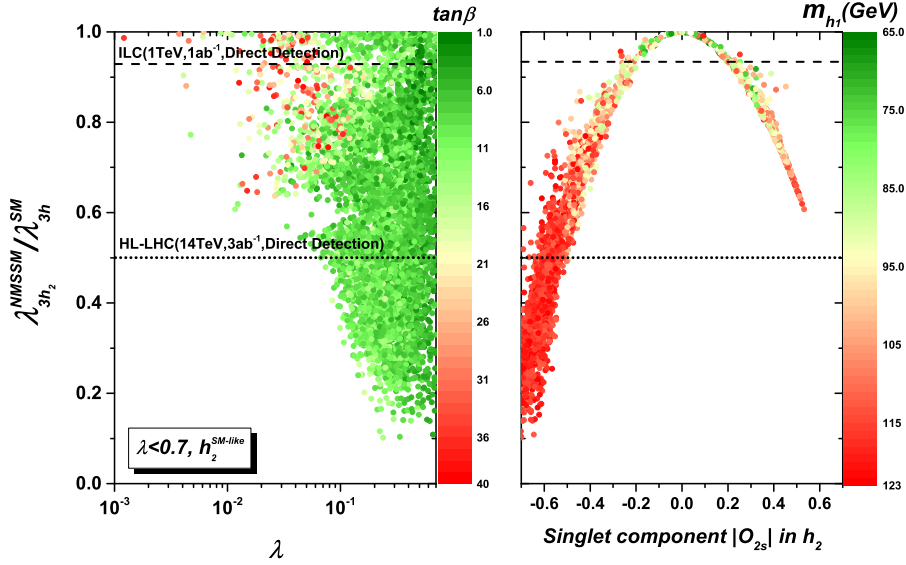}\vspace{-0.2cm}
\caption{Same as Fig.~\ref{nmssm-pars}, but showing the dependence of $\lambda^{NMSSM}_{3h_2}/\lambda^{SM}_{3h}$ versus $\lambda$ and the singlet component $|{\cal O}_{2s}|$ in $h_2$, where $m_{h_2} < 2 m_{h_1}$. The ILC(1 TeV, 1 ab$^{-1}$) and HL-LHC(14 TeV, 3 ab$^{-1}$) sensitivities are also plotted (the region below each horizontal line is detectable).}
\label{nmssm-3h}
\end{figure}
In Fig.~\ref{nmssm-3h}, we display the dependence of $\lambda^{NMSSM}_{3h_2}/\lambda^{SM}_{3h}$ versus $\lambda$ and $\tan\beta$ for the NMSSM ($\lambda<0.7$). As mentioned above, due to the Higgs mass constraint, most of the allowed model samples tend to have large values of $\lambda$. Furthermore, the ratio $\lambda^{NMSSM}_{3h_2}/\lambda^{SM}_{3h}$ is not sensitive to the value of $\tan\beta$. We also show the dependence of $\lambda^{NMSSM}_{3h_2}/\lambda^{SM}_{3h}$ versus the singlet component $|{\cal O}_{2s}|$ in the SM-like Higgs boson $h_2$ and the lightest CP-even Higgs mass $m_{h_1}$ for the NMSSM ($\lambda<0.7$). Besides, we plot the expected ILC(1 TeV, 1 ab$^{-1}$) and HL-LHC(14 TeV, 3 ab$^{-1}$) sensitivities to Higgs self-coupling from the direct measurements \cite{snowmass}. We can see that $\lambda^{NMSSM}_{3h_2}/\lambda^{SM}_{3h}$ becomes small with the increase of the singlet component $|{\cal O}_{2s}|$ and can minimally reach 0.1 in the allowed parameter space. Meanwhile, such a large mixing can make the mass of the lightest singlet-dominant CP-even Higgs boson $h_1$ close to the 125 GeV SM-like Higgs boson $h_2$.

\begin{figure}[ht]
\centering
\includegraphics[width=5in]{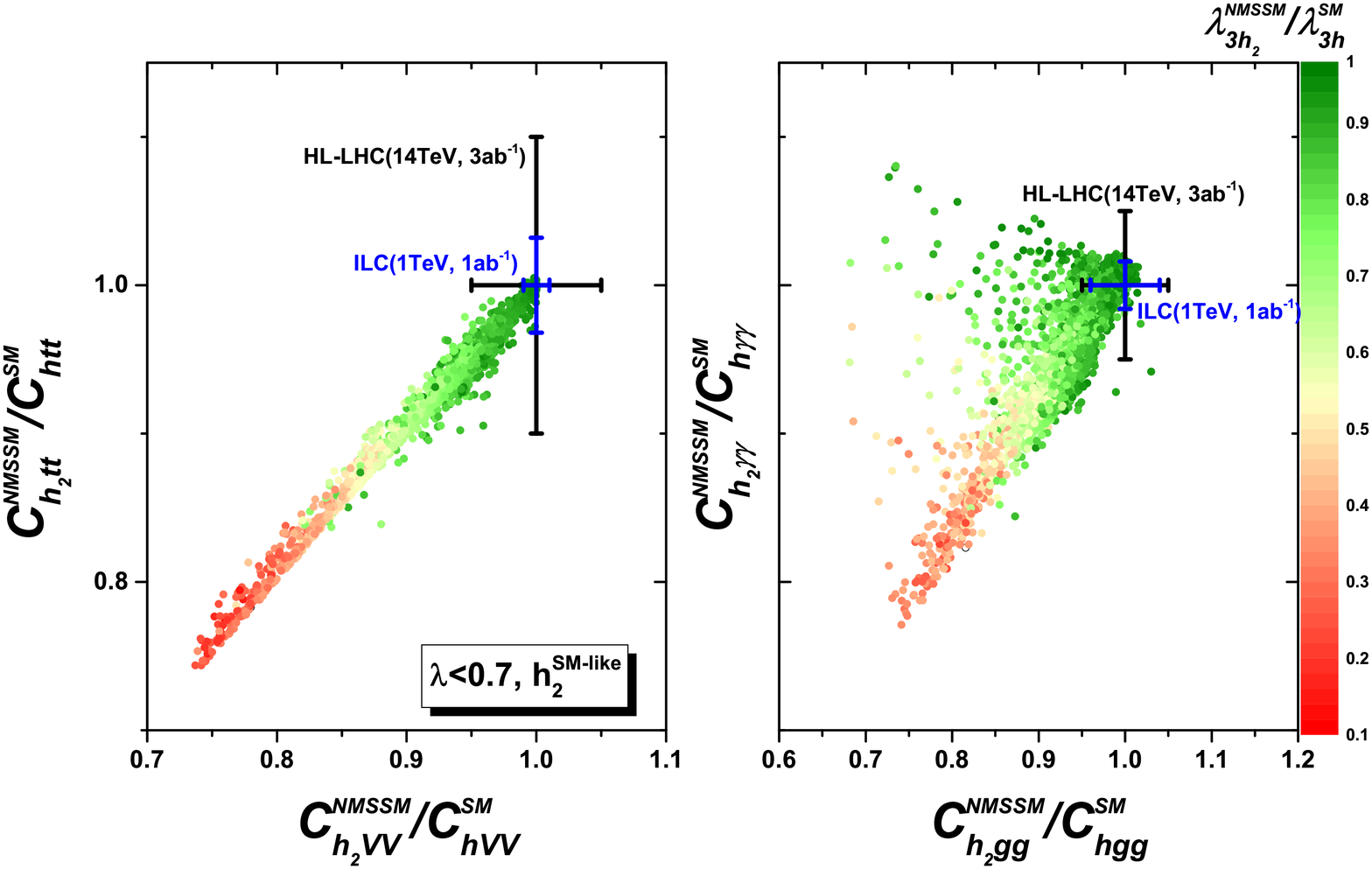}\vspace{-0.2cm}
\caption{Same as Fig.~\ref{mssm-couplings} but for the NMSSM ($\lambda<0.7$).}
\label{nmssm-couplings}
\end{figure}

\begin{table}
\caption{A benchmark point for the
NMSSM with $\lambda<0.7$ ($h_2$ is the SM-like Higgs boson). The cross sections (pb) are calculated for LHC-14 TeV.}
\begin{center}
\begin{tabular}{|c|c|c|c|c|c|c|c}
\hline
 $\lambda$ &$\kappa$ &$\tan \beta$ &$\mu$ (GeV) &$A_\lambda$ (GeV) &$A_\kappa$ (GeV) &$\lambda^{\rm NMSSM}_{3h_2}/\lambda^{\rm SM}_{3h_2}$\\
\hline
 0.16 &0.31 &17.9  &105.3  &1461.9  &-716.2 &0.798 \\
\hline \hline
$m_{h_1}$ (GeV)  &$\sigma_{ggh_1}$ &$\sigma_{VVh_1}$ &$\sigma_{Wh_1}$ &$\sigma_{Zh_1}$ &$\sigma_{t\bar{t}h_1}$ &$\sigma_{b\bar{b}h_1}$\\
\hline
119.56 &25.41 & 2.14 & 0.85 & 0.49 & 0.33 &0.54\\
\hline
$Br_{h_1\rightarrow\gamma\gamma}$ &$Br_{h_1\rightarrow gg}$ &$Br_{h_1\rightarrow ZZ^*}$ &$Br_{h_1\rightarrow WW^*}$ &$Br_{h_1\rightarrow c\bar{c}}$ &$Br_{h_1\rightarrow b\bar{b}}$ &$Br_{h_1\rightarrow\tau^+\tau^-}$\\
\hline
0.135\% &3.60\% &0.717\% &7.88\% &2.12\% &77.3\% &8.10\%\\
\hline \hline
$m_{h_2}$ (GeV)  &$\sigma_{ggh_2}$ &$\sigma_{VVh_2}$ &$\sigma_{Wh_2}$ &$\sigma_{Zh_2}$ &$\sigma_{t\bar{t}h_2}$ &$\sigma_{b\bar{b}h_2}$\\
\hline
127.30 &24.81 & 2.05 & 0.71 & 0.41 & 0.29  &0.14\\
\hline
$Br_{h_2\rightarrow\gamma\gamma}$ &$Br_{h_2\rightarrow gg}$ &$Br_{h_2\rightarrow ZZ^*}$ &$Br_{h_2\rightarrow WW^*}$ &$Br_{h_2\rightarrow c\bar{c}}$ &$Br_{h_2\rightarrow b\bar{b}}$ &$Br_{h_2\rightarrow\tau^+\tau^-}$\\
\hline
0.381\% &8.11\% &3.92\% &34.0\% &4.14\% &44.6\% &4.55\%\\
\hline
\end{tabular}
\end{center}
\label{nmssm-bp}
\end{table}
In Table \ref{nmssm-bp}, we present the properties of two CP-even Higgs bosons $h_1$ and $h_2$ for such a benchmark point at 14 TeV LHC. We can see that the cross sections of the single $h_1$ production are close to those of SM-like $h_2$ because of the large doublet and singlet mixing components in both $h_1$ and $h_2$. However, the branching ratio $h_1 \to VV (V=Z,W\gamma,g)$ is greatly suppressed by the increase of the partial width of $h_1 \to b\bar{b}$. Thus, the observed production rate of $gg \to h_1 \to VV$ is much smaller than that of SM-like $h_2$. We also checked and found that although the cross section of $gg \to h_1 \to \tau^+\tau^-$ can reach 2.06 pb, it is still smaller than the upper bound given by the current LHC searches for $H/A \to \tau^+\tau^-$ \cite{atlas-tautau}. Besides, it should be mentioned that the sizable modification of the self-coupling of the SM-like Higgs boson $h_2$ always accompanies with the great changes in other Higgs couplings, which can be seen in Fig.~\ref{nmssm-couplings}. So, given the limited sensitivity of measuring the Higgs boson self-couplings, we anticipate that the precision measurement of Higgs couplings to gauge bosons and fermions could test this scenario at the LHC and future colliders prior to the direct detection of triple or quartic Higgs couplings.

In Fig.~\ref{nmssm-couplings}, we show the Higgs couplings in the NMSSM ($\lambda<0.7$) surviving all the experimental constraints. For comparison, we also show the discovery potential of the expected ILC(1 TeV, 1 ab$^{-1}$) and HL-LHC(14 TeV, 3 ab$^{-1}$) \cite{snowmass}. From this figure, we obtain the following observations: (1) Due to the singlet admixture in the SM-like Higgs boson, both Higgs gauge couplings and top-Higgs coupling can be maximally reduced by about $30\%$ in the allowed region, which is much larger than in the MSSM. So, the expected measurements of the Higgs gauge couplings at the HL-LHC and ILC can exclude the parameter space with $\lambda^{NMSSM}_{3h_2}/\lambda^{SM}_{3h}<0.82$ and
$\lambda^{NMSSM}_{3h_2}/\lambda^{SM}_{3h}<0.93$, respectively; (2) With the increase of the singlet component in the Higgs couplings, both $C^{NMSSM}_{hgg}/C^{SM}_{hgg}$ and $C^{NMSSM}_{h\gamma\gamma}/C^{SM}_{h\gamma\gamma}$ are significantly reduced. On the other hand, due to the additional tree-level contribution ($\sim \lambda v \sin2\beta$) and the positive mixing effect, we find that a stop with mass less than 200 GeV is still allowed by the SM-like Higgs mass constraint in the NMSSM (similar results have been obtained in previous NMSSM works \cite{nmssm-higgs}). Consequently, the ratio $C^{NMSSM}_{hgg}/C^{SM}_{hgg}$ becomes larger than one, due to the constructive contribution from the light stop in loop. We also note that even when $\lambda^{NMSSM}_{3h_2}$ approaches to $\lambda^{SM}_{3h}$, $C^{NMSSM}_{h_2\gamma\gamma}/C^{SM}_{h\gamma\gamma}$ can still be enhanced by about 8\%.

\subsection{Results for the NMSSM ($\lambda>0.7$)}
\begin{figure}[ht]
\centering
\includegraphics[width=5in]{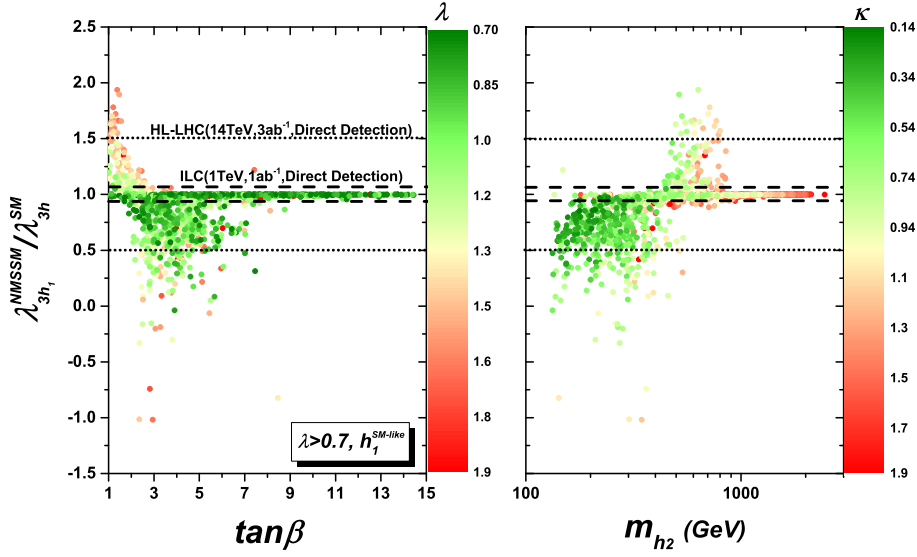}\vspace{-0.2cm}
\caption{The NMSSM ($\lambda>0.7$) samples surviving all the experimental constraints, showing the dependence of $\lambda^{NMSSM}_{3h_1}/\lambda^{SM}_{3h}$ versus $\tan\beta$ and $m_{h_2}$. The ILC(1 TeV, 1 ab$^{-1}$) and HL-LHC(14 TeV, 3 ab$^{-1}$) sensitivities are also plotted.}
\label{lambda-3h}
\end{figure}
In Fig.~\ref{lambda-3h}, we display the dependence of $\lambda^{NMSSM}_{3h_1}/\lambda^{SM}_{3h}$ versus $\tan\beta$ and $m_{h_2}$ for the NMSSM ($\lambda>0.7$).
Similar to Fig.~\ref{nmssm-pars}, the larger $\lambda$ is, the smaller $\tan\beta$ becomes to satisfy the requirement of the Higgs mass. The ratio $\lambda^{NMSSM}_{3h_1} / \lambda^{SM}_{3h}$ can vary from -1.1 to 1.9 in our scan. For example, the ratio $\lambda^{NMSSM}_{3h_1} / \lambda^{SM}_{3h}$ is equal to $1.89$ when $\lambda=1.51$ and $\kappa=0.67$, and is $-1.04$ for $\lambda=1.57$ and $\kappa=1.16$. Since we require $m_{h_3}>1$ TeV, for our samples with $m_{h_3}\gg m_{h_{1,2}}$, $\lambda^{NMSSM}_{3h_1} / \lambda^{SM}_{3h}$ is approximately proportional to $\lambda^2$ and becomes large with the increase of $\lambda$. In our scan ranges, such a feature could either enhance or suppress the Higgs self-coupling with respect to the SM prediction, and yield potentially large effects in Higgs pair production cross sections \cite{barbieri,cao}. However, in general, the two masses $m_{h_2}$ and $m_{h_3}$ are virtually independent and the mixing patterns are complicated. It is worth mentioning that the value 
of $\lambda^{NMSSM}_{3h_1} / \lambda^{SM}_{3h} \lessgtr 1$ strongly relies on the mass of the next-to-lightest CP-even Higgs boson $h_2$. To be specific, when $\kappa$ becomes small (large), $m_{h_2}$ is inclined to be light (heavy). If $m_{h_2}$ is lighter (heavier) than about 400 GeV, $\lambda^{NMSSM}_{3h_1} / \lambda^{SM}_{3h}$ is smaller (larger) than unity for most samples. The properties of $h_2$ will be discussed in the following. We should mention that the large $\lambda$ and $\kappa$ that produce the large deviation of $\lambda^{NMSSM}_{3h_1} / \lambda^{SM}_{3h}$ jeopardize the perturbativity up to GUT scale. Thus, the new unknown strong dynamics will appear at some cut-off scale $\Lambda$. On the other hand, the constructions of high scale theory by adding vector-like matter can allow for a larger $\lambda$ value and relax the cutoff scale to high values \cite{lambda-susy-extensions}, which is however beyond the scope of our study.

\begin{figure}[ht]
\centering
\includegraphics[width=4in,height=3in]{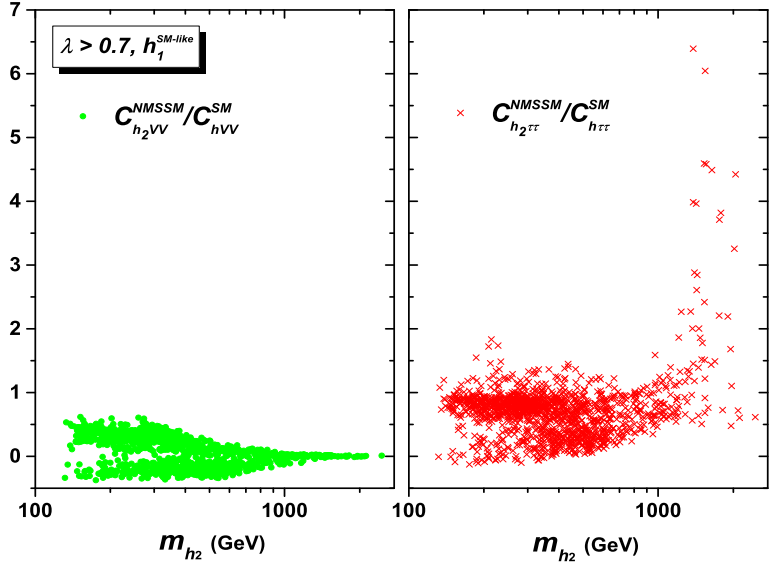}\vspace{-0.2cm}
\caption{The NMSSM ($\lambda>0.7$) samples surviving all the experimental constraints, showing the dependence of the Higgs couplings versus $m_{h_2}$.}
\label{lambda-mh2}
\end{figure}

\begin{table}
\caption{A benchmark point for the NMSSM with $\lambda >0.7$ ($h_1$ is the SM-like Higgs boson). The cross sections (pb) are calculated for LHC-14 TeV.}
\begin{center}
\begin{tabular}{|c|c|c|c|c|c|c|c}
\hline
 $\lambda$ &$\kappa$ &$\tan \beta$ &$\mu$ (GeV) &$A_\lambda$ (GeV) &$A_\kappa$ (GeV) &$\lambda^{\rm NMSSM}_{3h_2}/\lambda^{\rm SM}_{3h_2}$\\
\hline
 1.54 &0.75 &3.06  &474.00  &1035.41  &-644.30 &-0.203 \\
\hline \hline
$m_{h_1}$ (GeV)  &$\sigma_{ggh_1}$ &$\sigma_{VVh_1}$ &$\sigma_{Wh_1}$ &$\sigma_{Zh_1}$ &$\sigma_{t\bar{t}h_1}$ &$\sigma_{b\bar{b}h_1}$\\
\hline
126.9 &42.86 & 3.42 & 1.17 & 0.70 & 0.52 &0.29\\
\hline
$Br_{h_1\rightarrow\gamma\gamma}$ &$Br_{h_1\rightarrow gg}$ &$Br_{h_1\rightarrow ZZ^*}$ &$Br_{h_1\rightarrow WW^*}$ &$Br_{h_1\rightarrow c\bar{c}}$ &$Br_{h_1\rightarrow b\bar{b}}$ &$Br_{h_1\rightarrow\tau^+\tau^-}$\\
\hline
0.427\% &9.87\% &4.18\% &36.5\% &4.98\% &39.7\% &4.03\%\\
\hline \hline
$m_{h_2}$ (GeV)  &$\sigma_{ggh_2}$ &$\sigma_{VVh_2}$ &$\sigma_{Wh_2}$ &$\sigma_{Zh_2}$ &$\sigma_{t\bar{t}h_2}$ &$\sigma_{b\bar{b}h_2}$\\
\hline
282.0 & 6.33 & 0.90 & 0.06 & 0.03 & 0.03  &0.04\\
\hline
$Br_{h_2\rightarrow\gamma\gamma}$ &$Br_{h_2\rightarrow gg}$ &$Br_{h_2\rightarrow ZZ^*}$ &$Br_{h_2\rightarrow WW^*}$ &$Br_{h_2\rightarrow b\bar{b}}$ &$Br_{h_2\rightarrow\tau^+\tau^-}$ &$Br_{h_2\rightarrow h_1h_1}$\\
\hline
0.000663\% &0.0173\% &14.1\% &32.1\% &0.0996\% &0.012\% &53.6\%\\
\hline
\end{tabular}
\end{center}
\label{lambda-bp}
\end{table}

In Fig.~\ref{lambda-mh2}, we plot the couplings of $h_2$ with gauge bosons and $\tau$ leptons, normalized to the SM values. We can see that the gauge coupling $h_2VV$ is always suppressed due to the presence of singlet $(s)$ and non-SM doublet $(H)$ components in $h_2$. If $h_2$ is singlet-like, the $h_2\tau^+\tau^-$ coupling is suppressed as well, while if $h_2$ is non-SM doublet-like, $h_2\tau^+\tau^-$ coupling can be enhanced by $\tan\beta$. The detailed mixing patterns of $s$ and $H$ in $h_2$ and its couplings have been thoroughly investigated in \cite{cao}. We checked that the cross section $gg \to h_2 \to \tau^+\tau^-$ for our samples is at least one order lower than the current direct search bound on the non-SM Higgs bosons \cite{cms-heavy,atlas-tautau}. The main reason is that for a heavy Higgs boson $h_2$, the new decay channels, such as $h_2 \to h_1 h_1$ \cite{su}, can be opened and the branching ratio of $h_2 \to \tau^+\tau^+$ will be highly suppressed.

In Table~\ref{lambda-bp}, we present the properties of two CP-even Higgs bosons $h_1$ and $h_2$ for a benchmark point at 8 TeV LHC. We can see that all the SM branching ratios of $h_2$ are reduced due to the opened decay mode $h_2 \to h_1 h_1$, which can reach 53.6\% for $m_{h_2}=282.0$ GeV. Such a Higgs-to-Higgs decay will lead to a resonant SM-like di-Higgs bosons production $pp \to h_2 \to h_1 h_1$ at the LHC. A resonance feature in the $h_1 h_1$ invariant mass can be served as a smoking gun to search for the heavy Higgs boson $h_2$. With one SM-like Higgs boson $h_1$ decaying to two photons and the other decaying to $b$-quarks, the resonant signal may be observable above the di-Higgs continuum background for $m_{h_2} < 1$ TeV at the HL-LHC \cite{barger}.

\begin{figure}[ht]
\centering
\includegraphics[width=3.5in]{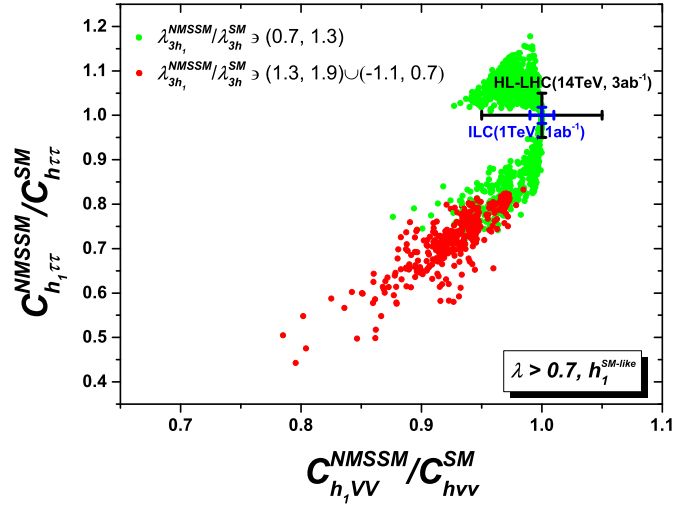}\vspace{-0.2cm}
\caption{Same as Fig.~\ref{lambda-mh2}, but showing the couplings of $h_1$. The HL-LHC(14 TeV, 3 ab$^{-1}$) and ILC(1 TeV, 1 ab$^{-1}$) sensitivities \cite{snowmass}
are also plotted.}
\label{lambda-couplings}
\end{figure}
Next, we present in Fig.~\ref{lambda-couplings} the couplings of the SM-like $h_1$ to weak gauge bosons and tau pair for the NMSSM ($\lambda>0.7$). To show their correlation with the Higgs self-coupling, we use the red color to highlight the points that satisfy $|\lambda^{NMSSM}_{3h_1}/\lambda^{SM}_{3h}-1|>0.3$. It can be seen that the large Higgs self-couplings corrections correspond to the sizable shifts in Higgs couplings with gauge bosons and tau pair. Since the tree-level mass of the SM-like Higgs boson $h_1$ could easily exceed 125 GeV, $h_1$ is likely to have non-negligible singlet and/or non-SM doublet components, which makes its couplings deviate from the SM predictions. From this figure, it can be seen that $C^{NMSSM}_{h_1VV}/C^{SM}_{hVV}$ is always less than unity since any singlet and/or non-SM doublet components in $h_1$ will make the Higgs couplings to weak gauge bosons smaller than the SM predictions. However, as mentioned above, if the next-to-dominant component in $h_1$ is the non-SM doublet, the Higgs coupling with the down-type fermions may be enhanced and larger than the SM predictions, such as the $h_1 \tau^+ \tau^-$ coupling. Therefore, the future measurements of $C_{h_1\tau^+\tau^-}$ and $C_{h_1VV}$ couplings can give strong constraints on the parameters space of the NMSSM with $\lambda>0.7$ and set limits on the Higgs self-coupling $C^{NMSSM}_{3h_1}$.

From the above discussions, we can see that the large deviation of the Higgs self-coupling $\lambda_{3h_1}$ for the NMSSM ($\lambda>0.7$) is always accompanied by other collider signatures, such as a shift in the Higgs couplings and the production on resonance of the non-SM doublet $h_2$. At the LHC, the resonant production of $h_2$ may be observed through the channels $gg \to h_2 \to \tau^+\tau^-,h_1h_1$. However, the sensitivities of these channels strongly depend on the mass of $h_2$ and on its singlet and non-SM doublet components. If $h_2$ is dominantly singlet, both direct searches will not be powerful in probing our scenario at the LHC since all the $h_2$ couplings to SM particles are greatly reduced. Thus, the precision measurement of the observed 125 GeV Higgs boson couplings will play an unique role in probing this scenario at future colliders. If $h_2$ is dominantly doublet (non-SM), and if $m_{h_1}<m_{h_2}<2m_{h_1}$, the process $gg \to h_2 \to \tau^+\tau^-$ is still greatly suppressed due to the reduction of the coupling $h_2 t\bar{t}$ despite the fact that the coupling $h_2\tau^+\tau^-$ can be maximally enhanced by a factor of 2. On the other hand for $2m_{h_1}<m_{h_2}$, the decay $h_2 \to h_1 h_1$ is open and contributes to the cross section of $pp \to h_1h_1$. Therefore, besides the Higgs coupling measurements, a resonance feature in the $h_1 h_1$ invariant mass or an excess in the inclusive $h_1 h_1$ production can be used to probe this model at the future LHC.

\section{Conclusion\label{section4}}
We examined the currently allowed values of trilinear self-couplings of the SM-like 125 GeV Higgs boson ($h$) in the 
MSSM and NMSSM after the LHC Run-1. Considering all the relevant experimental constraints, such as the Higgs data, the flavor constraints, the electroweak precision observables as well as the dark matter detections, we performed a scan over the parameter space of each model and obtained the following observations:
\begin{itemize}
\item In the MSSM, the Higgs self-coupling  is suppressed relative to the SM value. Such a suppression was found to be rather weak and the ratio $\lambda^{\rm MSSM}_{hhh}/\lambda^{\rm SM}_{hhh}$ is above 0.97 due to the tightly constrained parameter space, cf. Figs.~\ref{mssm-low} and \ref{mssm-3h};
\item In the NMSSM with $\lambda<0.7$, we consider the case that the SM-like Higgs boson mass $m_{h_2}$ is less than twice of $m_{h_1}$, so that $h_2$ will not decay into a pair of $h_1$. We found that the Higgs self-coupling was found to be likely suppressed and the ratio $\lambda^{\rm NMSSM}_{hhh}/\lambda^{\rm SM}_{hhh}$ can be as low as 0.1 due to the large mixing between singlet and doublet Higgs bosons, cf. Fig.~\ref{nmssm-3h}. In that case, the coupling of the SM-like Higgs boson ($h_2$) to to $W$ and $Z$ bosons and top quark pairs are all suppressed as compared to the SM prediction. On the other hand, its couplings to loop-induced processes, such as photon pairs or gluon pairs, can be enhanced, cf. Fig.~\ref{nmssm-couplings}. Given the limited sensitivity of measuring the Higgs boson self-couplings, we anticipate that the precision measurement of Higgs couplings to gauge bosons and fermions could test this scenario at the LHC and future colliders prior to the direct detection of triple or quartic Higgs couplings;
\item In the NMSSM with $\lambda>0.7$ (also called $\lambda$-SUSY), the Higgs self-coupling can be greatly suppressed or enhanced relative to the SM value (the ratio $\lambda^{\rm NMSSM}_{hhh}/\lambda^{\rm SM}_{hhh}$ can vary from -1.1 to 1.9), cf. Fig.~\ref{lambda-3h}, when $h_1$ is taken as the SM-like Higgs boson. Its coupling to $W$ and $Z$ bosons are always suppressed as compared to the SM predictions. On the contrary, its couplings to tau pairs can be either enhanced or suppressed relative to the SM value, cf. Fig.~\ref{lambda-couplings}. When $m_{h_2} > 2 m_{h_1}$, it is possible to observe a new resonance production in the di-Higgs boson channel at the LHC. While the couplings of $h_2$ to $W$ and $Z$ bosons are always suppressed, its coupling to tau pair can deviate largely from the SM value, depending on the mass of $h_2$, cf. Fig.~\ref{lambda-mh2}.
\end{itemize}
Since the NMSSM can give rather different values (compared with the SM) for the trilinear self-couplings of the SM-like Higgs boson, the future collider experiments like the high luminosity LHC or ILC can probe NMSSM through measuring the Higgs self-couplings.

\section*{Acknowledgement}
Lei Wu thanks Roman Nevzorov, Ulrich Ellwanger, Archil Kobakhidze and Michael Schmidt for very helpful discussions. This work was partly supported by the Australian Research Council, by the CAS Center for Excellence in Particle Physics (CCEPP), by the National Natural Science Foundation of China (NNSFC) under grants Nos. 11305049, 11275057, 11405047, 11275245, 10821504 and 11135003, by Specialized Research Fund for the Doctoral Program of Higher Education under Grant No.20134104120002, and by the U.S. National Science Foundation under Grant No. PHY-1417326.

\end{document}